# Event-Based Dynamic Banking Network Exploration for Economic Anomaly Detection

Andry Alamsyah, Dian Puteri Ramadhani, Farida Titik Kristanti

School of Economics and Business, Telkom University, Indonesia

andrya@telkomuniversity.ac.id

**ABSTRACT**

The instability of financial system issues might trigger a bank failure, evoke spillovers, and generate contagion effects which negatively impacted the financial system, ultimately on the economy. This phenomenon is the result of the highly interconnected banking transaction. The banking transactions network is considered as a financial architecture backbone. The strong interconnectedness between banks escalates contagion disruption spreading over the banking network and trigger the entire system collapse. This far, the financial instability is generally detected using macro approach mainly the uncontrolled transaction deficits amount and unpaid foreign debt. This research proposes financial instability detection in another point of view, through the macro view where the banking network structure are explored globally and micro view where focuses on the detailed network patterns called motif. Network triadic motif patterns used as a denomination to detect financial instability. The most related network triadic motif changes related to the instability period are determined as detector. We explore the banking network behavior under financial instability phenomenon along with the major religious event in Indonesia, Eid al-Fitr. We discover one motif pattern as the financial instability underlying detector. This research help to support the financial system stability supervision.

## 1. INTRODUCTION

The global financial turmoil in 2008 has spread destructive impact among the banks in a short time [1]. The financial system instability due to bank liquidity problems might trigger bank failures then generate spillover and contagion effects which have negative impact on the financial system, ultimately on the economy [2]. Rapid effect dissemination indicates that the entities within the interbank financial markets are highly interdependent to each other in a large banking network [1]. Thus, discovering the interbank financial market activity through the overall system approach becomes more important than the individual bank approach [1]. The defect spreading driven by the highly interconnected banking network [3] [4] trigger the entire system to collapse.

Financial instability prominently reduces the volume of commodity trading e.g., income taxes, personal income, and profits. This declination leads to potential disruptions in any sector that closely related to the welfare of human life [5]. The obstruction of human well-being driven by this contagion disruptions phenomenon has attracted many researchers to gain benefit through the network approach.

Network has been known as a powerful approach to represent a complex system by explaining entities interconnectedness. The network concept is suitable to be adopted in many real-world domains to gain the complex interactions knowledge. A network is basically constructed upon a set of nodes and edges between them. The banking network, as a network of transaction, is made of banks as nodes and fund transfers from a bank to other bank as edges. As the banking network is highly interconnected, when an important entity fails, the entire network is facing the collapse risk through the defaults cascade [6].

A network is usually characterized by its macro and the micro network structures. The macro network structure characterized by the network properties measurement while the micro network structures characterized by the network motifs [7]. Several researches have been conducted network characterization in various fields e.g.

communication, information dissemination, company networks, biological and non-biological network [7] [8] [9]. In the financial domain, network approach is accomplished to mine valuable insight related to the certain situation through identifying banks transacting behavior in various situations [1].

Many studies have been devoted to discovering the dynamic of banking network structure responding to various situations. For example, the bank connectivity within payment system is decline sharply after the September 11, 2001 attack. After the tragedy, payment activity rises above the daily average. Major religious event such as Thanksgiving, Good Friday, and Christmas Eve are also known as the reason of the banking network structure shift [1] [10] [11]. In Indonesia, the major religious event is Eid al-Fitr, which is the religious event of Muslims, the majority religion in Indonesia. There are more than 207 million Muslims in Indonesia or about 87.2% of the total population [12]. During this event, millions of workers and professionals returned from the metropolitan cities and brought billions of moneys to their hometowns to spend with families. This event might moderate the money circulation disruption [20].

Network motifs decompose a complex network into certain small-connected nodes patterns. Each of the patterns possess a set of unique characteristics [9]. The time-series motif trend changes along with the major religious event period provides a better understanding of the most changing motif pattern as an early warning of financial symptoms. This scenario might prevent the wide financial instability risks to trigger the economic system collapse.

Financial instability detection is substantial to prevent further wide loss spreading. This study offers financial instability detection through the micro approach. Based on the previous study [13] [14] [21], the network motifs are used as a denomination to detect the financial instability by observing the microscopic patterns dynamic within a network. In detecting financial instability, a detector is needed. Thus, a question emerges, how to determine the detector? We investigate the most representable motif pattern as a financial instability detector in Indonesia. To the best of our knowledge, this scenario has never been conducted in Indonesia. The state of the art in predicting financial instability through network motifs is introduced by Squartini [21] which implemented to Dutch banking network. We enhance his idea to Indonesia banking network case study, which has different transactions characteristics. We also introduced macro network measurement to enrich the methodology. The findings contribute to develop a financial instability detection model using macro and micro network approach for Indonesia.

This study presents an empirical analysis of the even-based dynamic banking network exploration in Indonesia. The religious event of the majority population in Indonesia is chosen to model financial instability scenario. We measure the banking transacting behavior responding to the event. Network properties measurement is used to understand the overall system's behavior comprehensively, while network motifs discovery is used to understand the microscopic system's behavior. Furthermore, we identify which motif(s) that become best financial instability detector. We explore the triadic motifs dynamic by calculating each of 16 patterns appearance in daily time slices. Finally, we observe the relation between the macro and micro network structure changes of a banking network and the event period. Based on this background, we came up with questions: which of the macro network structure that change along with the event? Which of the triadic motif pattern(s) reveal(s) the major micro network structural changes along with the event?

## 2. LITERATURE REVIEW
### 2.1 Network Analysis
Network study has developed over a decade. Currently, the application of network studies has existed in many fields, such as computer science, biology, and economics. The network consists of set of nodes which are connected by edges [7]. The nodes are illustrating actors or organizations, and edges are symbolizing interaction across the nodes. Edges states the actor's relations [15] [16] [17] or connections between actors [18] [19]. According to Barabási [23], the edge's direction expresses the interaction types, whether directed or undirected. The edges also have a weight, which indicates the value of each node's interaction [23].

Researchers have utilized network analysis to understand the real complex problem. The network approach emphasizes two main perspectives i.e., relationships between the entities and network structural form [28]. The network represents a complex system by clarifying the entity's linkage [21]. Hence, this approach is powerful in discovering the real system behavior of the complex problem.

There are several sub-sections in network science, one of which is a complex network. The complex network is a network with a non-regular topological property. By analyzing the complex network, we are able to draw knowledge over the complexity of structural, entities, and relation that occurred in the network. The structural form of complex networks also classified into different categories, such as random network, clustered random network, small-world network, core-periphery network, scale-free network, etc. [21].

Dynamic Network Analysis (DNA) is the ingenious approach that emerges trough the limitation of ordinary network analysis. DNA is also known as network temporal analysis. However, the temporal analysis is not always correlated with DNA since this approach is affected by several independent external factors that are not tied in network features. Temporal network analysis is more concern about revamping a network into a static time based on a fixed time interval [6] [37] [38].

DNA is dynamic, more extensive, complex, and likely have more suspicious factor compared to a classical network. DNA provides techniques which able to measure dynamic behavior on the complex system [23] [39]. The purpose of DNA is to discover the entity's behavior linked to how these entities affect the structural form on a network over a period of time [24]. The result of DNA provides valuable information regarding the pattern of entities in the complex problem.

**2.2 Network Properties Measurement**

The network has several properties to help us gain a better perspective of the complex system. Based on our research objectives, the network topological properties or known as network properties measurement is appropriate to be implemented for describing network structure [40] [41]. The relations between nodes within a network reveal the structure of a network [42] [46]. The several well-known network properties measurement is shown in Table 1.

*Table 1: Network Properties Measurement*

| Network Topology | Explanation |
|---|---|
| Number of Nodes | Number of entities in the network, refer to the network size [24] |
| Number of Edges | Number of interactions among entities [24] |
| Average Distance | Average distance between any paired entities in the network [8] |
| Network Density | Ratio of existing interaction among entities compared to potential relation in the network [8] |

**2.3 Network Motif**

The complex network has a diverse pattern of entities regarding its complexity. Network motifs are known as an imperative approach to represent a broad range of the natural system in the network [21]. This approach is significantly discovering patterns among entities interconnections. Network motif discovery is a complex approach. The motif is defined as the sub-graph networks, which are appeared regularly in the network, specifically in the directed network [33].

The network motifs have several motif types. It is classified based on the number of entities involved, like dyadic and triadic motifs. The dyadic motif is composed by two entities, while the triadic motif is established through three entities. The triadic network motif consists of more sufficient number of patterns combination than dyadic motif [31] [32] [33]. The triadic motif has 16 pattern combinations of entities relations shown in figure 1.

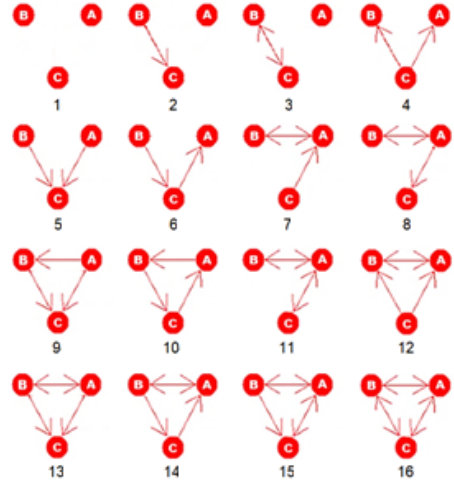

*Figure 1: 16 Patterns Combination of Triadic Motif*

Each pattern of triadic motifs represents different meanings to indicate the alteration of whole network structure. The points on triadic motifs are represent the nodes while the edge denote the interaction's direction. These patterns have disparate meaning depend on the knowledge domain.

### 2.4 Banking Network

Bank is a financial institution deals with credits and debits [43]. Banks play an important role in driving the economic growth of a country. Various business sectors need banks as their business development partners. Banks are defined as financial institutions whose main function is to collect funds to the public and provide banking services [25].

In a network, major nodes are possessed low connectivity and there are only a few strongly interconnected nodes. The banking network is built up with well-connected credit transactions between banks. It is constructed as a directed and weighted network. This network is composed of banks as a node or entity and financial transactions as edges [44] [45]. The edge's direction in this network represents the flow of banking transactions from the sender and receiver. The real-world banking system build up upon disparate value of transactions. Hence, we state the value of transactions as a weight of any node transactions. The banking network is susceptible to internal and external disturbances. When one of the entities is collapsed, the whole network is potentially exposed by the similar cascade collapse risk [34] [35] [36].

## 3. RESEARCH METHODOLOGY

We obtain the Indonesian banking transaction from the government financial agency. The banking transaction data are collected from January 2006 to December 2015. This period is passes through Eid al-Fitr period for 10 years. The dataset is not publicly available since comprising credential information. We extract the components relationships within the dataset by masking the sensitive information to maintain the confidentiality. This research workflow is shown in figure 2.

The banking transaction data is constructed upon 11.007.985 rows and 9 features. We reduce the data dimensionality by choosing only the most appropriate features to gain the desired result. In gaining knowledge through large scale data to solve real-world problems in more secrecy way, data sensitivity becomes the major challenge. We solve this challenge by masking the dataset. The banking transaction dataset is masked through data alteration to conserve the confidentiality and anonymity. We arrange the masking process by retaining the original data distribution to equate the proportion of transaction value. Through this process, we generate 120 months masked banking transaction data starting from January 2006 to December 2015. The selected features included transaction date, origin bank, destination bank, and the number of transactions as showed in table 2. We store the transaction data in daily time slices and mask each bank identity record. This masked dataset allows us to obtain desired knowledge without inflicting significant privacy and security concerns.

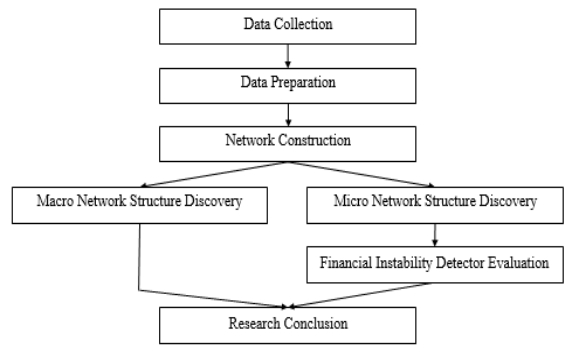

*Figure 2: Research Workflow*

*Table 2: Banking Transaction Dataset Example*

| Transaction Time | Masked Origin Bank | Masked Destination Bank | Number of Transaction(s) |
|---|---|---|---|
| 10 / 5 / 2006 | A | B | 1 |
| 10 / 5 / 2006 | C | B | 25 |
| 10 / 7 / 2006 | B | D | 7 |
| 10 / 8 / 2006 | D | A | 71 |

The final transaction data is modeled into banking networks. A banking network is built up upon banks as nodes and the banking transactions as edges. We modeled the network as a weighted and directed graph. The edge weight states the transaction value while edge direction shows transaction flow from the origin bank as fund sender to the destination bank as fund receiver.

The macrostructure of daily banking networks is discovered through network properties measurement. We recapitulate the findings by deploying each of the daily banking network properties measurement result as a time-series graph. The fluctuation of each network properties then analyzed to map the change along with the instability period. Therefore, the macro view of banking transaction behavior under instability is identified completely.

The microstructure of daily banking networks is discovered through network motifs. We observe each network triadic motif from 16 patterns combination within the daily banking network. We recapitulate the findings by deploying each of the 3-node sub-graphs discovery research as a time-series graph. Through this graph, the trend is observed to map out each 3-node sub-graphs change under the financial instability due to the religious event. Therefore, the detail view of banking transaction behavior under financial instability is identified completely. After discovering the dynamic of each triadic motif pattern under financial instability, then we identify the most dominance pattern(s) to signify the banks' behavioral changes to be used as the detector of financial instability condition.

## 4. RESULT AND DISCUSSION

This study highlights the major fluctuation trend changes in the macro and micro network structure related to Eid al-Fitr event. We construct the whole and daily banking transaction networks from January 2006 to December 2015 period. This period is passes through 11 Eid al-Fitr period as showed in table 3. After constructing the banking network, we focus on the recent 2 years banking network dynamic (event in 2014 and 2015) to obtain deeper knowledge.

*Table 3: Religious Event Dates*

| Year | Event Date |
|---|---|
| 2006 | 24th – 25th October |
| 2007 | 13th – 14th October |
| 2008 | 1st – 2nd October |
| 2009 | 21st – 22nd September |
| 2010 | 10th – 11th September |
| 2011 | 30th – 31st August |
| 2012 | 19th – 20th August |
| 2013 | 8th – 9 th August |
| 2014 | 28th – 29th July |
| 2015 | 17th – 18th July |

From the 120 months banking transaction data, we discover that the overall banking network is a scale free network [28] [29] characterized by having power-law weighted degree distribution [30]. Power-law distribution indicates there are only a few banks that having many transaction activities while there are many banks that having only a few transaction activities. This is similar to core-periphery bank structure which is already well-known in banking studies. The whole banking network weighted degree distribution is shown in figure 3. The daily banking network constructed upon daily banking transaction data in 120 months, or we call the macro network is visualized in figure 4. Darker node color indicates the more well-connected bank.

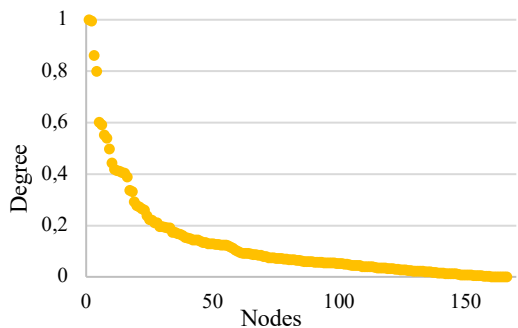

*Figure 3: The 120 Months Banking Network Degree Distribution*

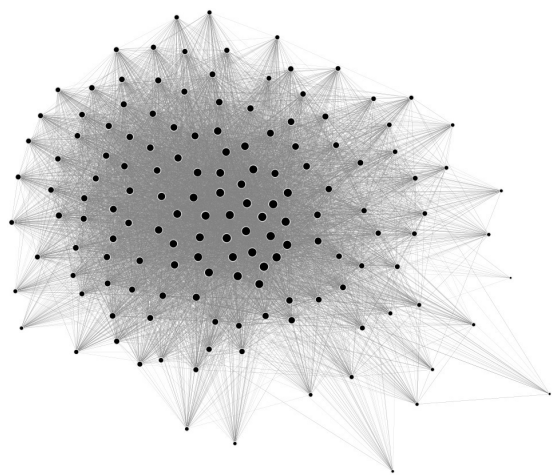

*Figure 4: The 120 Months Banking Network Visualization*

We investigate the macro network over 120 months and observe comprehensively in the recent 2 year's event (2014 – 2015). The observation focuses on the duration of 15 days before and after the event in 2014 and 2015. We use the network properties measurement in daily time slices to discover the banks transacting behavior in macro view and the network triadic motif combination in daily time slices to discover the bank transacting behavior in micro view.

Z-score is used to show fluctuations of daily measurement fluctuation. Z-score compare the daily measurement results to the average population in 10 years. Z-score is a means to reveal measurement deviations given from the population [26]. This score illustrates how far the measurement deviation from the average population measurement [27]. Z-scores formula are calculated as in

$$Z = (X - \mu) / \sigma \quad (1)$$

Where (X) denote the observed value, (μ) denote the population average value, and (σ) denote the population deviation standards.

Z-scores above the population average are stated by a positive value, while a Z-score below the population average stated by a negative value. The greater of the Z-score value (in the positive or negative direction), the greater the magnitude of the deviation from the average [27].

### 4.1 Macro Network Structure Discovery

We show the dynamic of network properties from overall time span, but we seek to explain from only the 2014-205 event, to improve the clarity of the study to the most detail event. Figure 6 shows the dynamic number of nodes, which fluctuate between 135 and 147 with the average of 143. We discover that the number of nodes fluctuate around normal line but having major declination in the day+1. After the declination, number of nodes increased and fluctuate normally.

Figure 8 shows the dynamic number of edges which fluctuate between 1180 and 6164 with the average of 4812,4. We discover that the number of edges is fluctuating around normal line and only having sharp declination in the day+1. In the day+2, the number of edges already increased and back to fluctuate around normal line.

Figure 10 shows the dynamic measurement of average distance. They fluctuate between 1.79 and 2.36 with the average of 1.87. We discover that the average distance is having minor fluctuation around normal line. Figure 12 shows the dynamic measurement of network density. They fluctuate between 0.07 and 0.30 with the average of 0.24. We discover network density only declining in the day+1. The next day after the declination, network density are fluctuating normally.

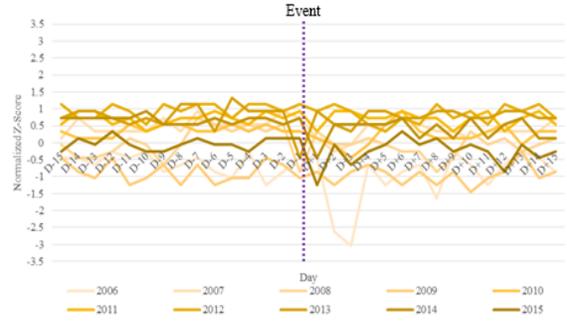

*Figure 5: The Dynamic of Number of Nodes (2006 – 2015)*

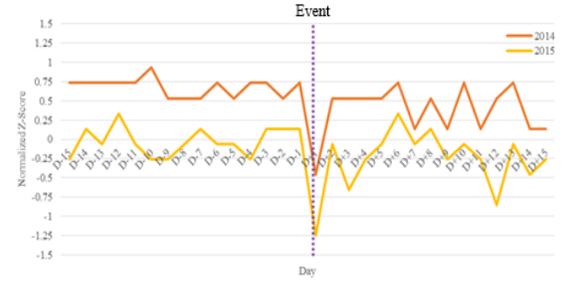

*Figure 6: The Dynamic of Number of Nodes (2014 – 2015)*

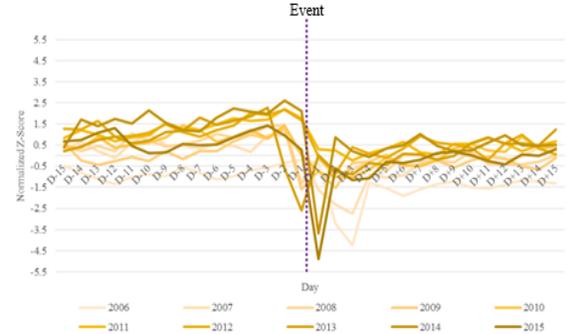

*Figure 7: The Dynamic of Number of Edges (2006 – 2015)*

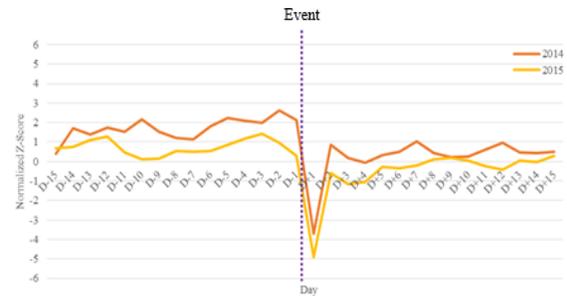

*Figure 8: The Dynamic of Number of Edges (2014 – 2015)*

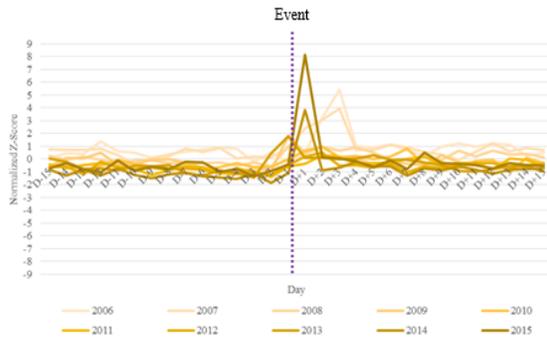

*Figure 9: The Dynamic of Distance (2006 – 2015)*

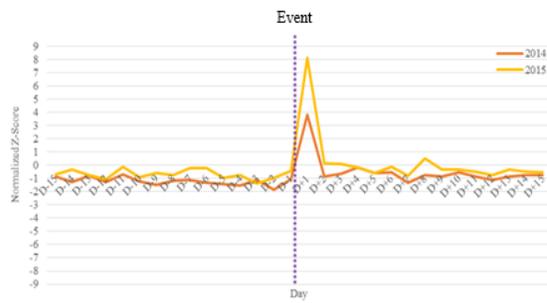

*Figure 10: The Dynamic of Distance (2014 – 2015)*

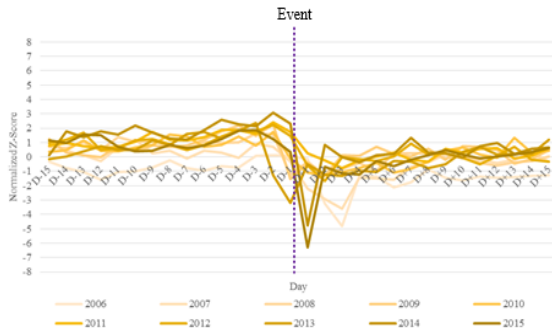

*Figure 11: The Dynamic of Network Density (2006 – 2015)*

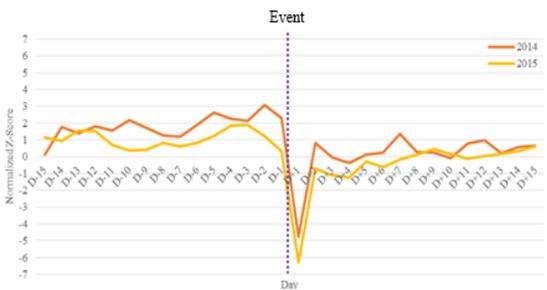

*Figure 12: The Dynamic of Network Density (2014 – 2015)*

## 4.2 Micro Network Structure Discovery

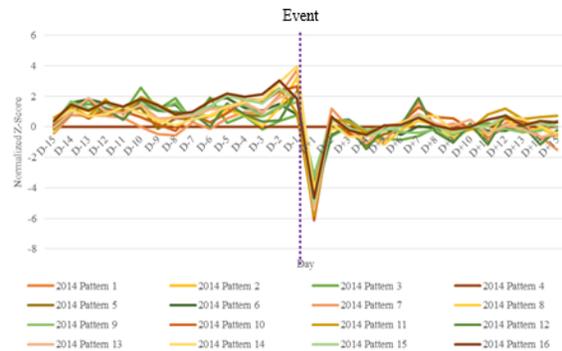

*Figure 13: The Dynamic of Triadic Motif (2014)*

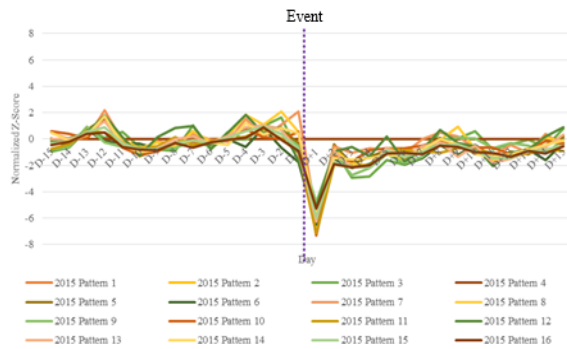

*Figure 14: The Dynamic of Triadic Motif (2015)*

Figure 13 and figure 14 show the fluctuation of each triadic motif pattern 15 days before and after 2014 and 2015 event. All of the triadic network motif patterns show the similar dynamic trend. In day-15 to day-3 of each year event, the patterns are fluctuating normally without forming certain trend. The trend shift begins since the day-2. Some of the patterns are increased in from the day-2 to day-1. After the increasement, all of the triadic motif patterns are decreased sharply in one day after each year event. In the following day, triadic motif patterns are recovered in normal fluctuation.

Based on the results, we found that pattern 16 is having the major changes and the most correlating triadic motif pattern to detect early warning of the financial instability. In 2014, pattern 16 decline 87% from the initial score in day+1 and increase 564% in day+2. Meanwhile, in 2015, pattern 16 decline 93% from the initial score in day+1 and increase 990% in day+2.

## 4.3 Financial Instability Detector Evaluation

Table 4. Daily Motif Patterns Change 15 Days Before and After the 2014 Event

| Day | P1 | P2 | P3 | P4 | P5 | P6 | P7 | P8 | P9 | P10 | P11 | P12 | P13 | P14 | P15 | P16 |
|---|---|---|---|---|---|---|---|---|---|---|---|---|---|---|---|---|
| D-15 | 0% | 0% | 0% | 0% | 0% | 0% | 0% | 0% | 0% | 0% | 0% | 0% | 0% | 0% | 0% | 0% |
| D-14 | 0% | 0% | 11% | 0% | 12% | 14% | 13% | 21% | 22% | 11% | 8% | 16% | 13% | 11% | 17% | 18% |
| D-13 | 0% | 0% | 1% | 0% | 2% | 2% | -1% | -3% | -6% | -6% | -7% | 12% | 11% | -5% | -3% | -5% |
| D-12 | 0% | 0% | -2% | 0% | -8% | -2% | 0% | -2% | -4% | 6% | 13% | -8% | -9% | 2% | 5% | 8% |
| D-11 | 0% | 0% | -2% | 0% | 2% | -4% | -1% | 2% | 5% | -1% | -10% | -8% | -2% | 7% | -3% | -4% |
| D-10 | 0% | 0% | 13% | 0% | 2% | 7% | -6% | 5% | 2% | -4% | 14% | 18% | 6% | -5% | 7% | 7% |
| D-9 | 0% | 0% | -10% | 0% | -11% | -5% | -5% | -14% | -5% | -5% | -5% | -8% | -10% | -6% | -7% | -6% |
| D-8 | 0% | 0% | 6% | 0% | 6% | 2% | -1% | 7% | 6% | -4% | -6% | -1% | 1% | -5% | -6% | -9% |
| D-7 | 0% | 0% | 2% | 0% | 9% | 7% | 15% | 8% | 0% | 11% | 7% | 27% | 9% | 5% | 11% | 9% |
| D-6 | 0% | 0% | 12% | 0% | 2% | 14% | -5% | -2% | 12% | -5% | 10% | -12% | 5% | 4% | 7% | 11% |
| D-5 | 0% | 0% | -12% | 0% | 4% | 1% | 8% | 5% | 0% | 6% | 3% | 24% | 1% | 8% | 8% | 7% |
| D-4 | 0% | 0% | 4% | 0% | -3% | -6% | 4% | -3% | 2% | 9% | -3% | -9% | 4% | 3% | -3% | -3% |
| D-3 | 0% | 0% | -7% | 0% | -1% | -1% | -1% | -8% | -9% | -1% | 0% | -6% | -6% | 2% | -3% | 2% |
| D-2 | 0% | 0% | 5% | 0% | 4% | 4% | 12% | 15% | 8% | 7% | 8% | 2% | 11% | 13% | 12% | 12% |
| D-1 | 0% | 0% | 3% | 0% | 9% | -7% | 14% | 21% | 4% | 3% | -16% | 27% | 3% | 11% | -6% | -14% |
| D+1 | 0% | 0% | -33% | 0% | -53% | -51% | -60% | -65% | -58% | -73% | -77% | -81% | -78% | -79% | -81% | -87% |
| D+2 | 0% | 0% | 35% | 0% | 81% | 95% | 101% | 101% | 104% | 206% | 314% | 258% | 270% | 215% | 330% | 564% |
| D+3 | 0% | 0% | 5% | 0% | -4% | 1% | -11% | -1% | -1% | -12% | -3% | 9% | -8% | -13% | -7% | -14% |
| D+4 | 0% | 0% | -1% | 0% | -8% | -8% | -9% | -3% | -9% | 0% | -9% | -23% | -13% | 0% | -8% | -6% |
| D+5 | 0% | 0% | -5% | 0% | 7% | 2% | 0% | -10% | 4% | 5% | 4% | 28% | 17% | 6% | 7% | 12% |
| D+6 | 0% | 0% | -1% | 0% | 2% | -5% | 12% | 14% | -6% | 3% | 4% | 3% | 2% | 9% | 3% | 1% |
| D+7 | 0% | 0% | 2% | 0% | 9% | 7% | 15% | 8% | 0% | 11% | 7% | 27% | 9% | 5% | 11% | 9% |
| D+8 | 0% | 0% | 5% | 0% | -5% | -2% | -14% | -3% | 3% | -6% | -7% | -27% | -8% | 1% | -9% | -9% |
| D+9 | 0% | 0% | -4% | 0% | -5% | -4% | 2% | -11% | -12% | -1% | -8% | -10% | -2% | -13% | -7% | -4% |
| D+10 | 0% | 0% | 4% | 0% | 5% | 7% | -3% | 8% | 17% | -7% | 4% | 22% | 5% | 5% | 7% | 3% |
| D+11 | 0% | 0% | 2% | 0% | -8% | -4% | -8% | -4% | -6% | 3% | 13% | -20% | -11% | -1% | 1% | 8% |
| D+12 | 0% | 0% | -2% | 0% | 7% | 0% | 12% | 14% | 4% | 5% | 5% | 35% | 7% | 12% | 11% | 5% |
| D+13 | 0% | 0% | -2% | 0% | 0% | 4% | 0% | -11% | -3% | -2% | -8% | -9% | -3% | -3% | -9% | -10% |
| D+14 | 0% | 0% | 0% | 0% | -7% | 1% | -9% | 4% | 1% | -2% | 1% | -19% | -7% | -7% | 2% | 5% |
| D+15 | 0% | 0% | 5% | 0% | 3% | 2% | -10% | -7% | 10% | -7% | 1% | 17% | 3% | -7% | -2% | -2% |

Table 5. Daily Motif Patterns Change 15 Days Before and After the 2015 Event

| Day | P1 | P2 | P3 | P4 | P5 | P6 | P7 | P8 | P9 | P10 | P11 | P12 | P13 | P14 | P15 | P16 |
|---|---|---|---|---|---|---|---|---|---|---|---|---|---|---|---|---|
| D-15 | 0% | 0% | 0% | 0% | 0% | 0% | 0% | 0% | 0% | 0% | 0% | 0% | 0% | 0% | 0% | 0% |
| D-14 | 0% | 0% | 3% | 0% | -6% | 6% | -5% | -1% | 3% | -2% | 5% | 3% | -1% | -8% | 1% | 5% |
| D-13 | 0% | 0% | 12% | 0% | 12% | 12% | 1% | 8% | 25% | -4% | 14% | 7% | 6% | 6% | 12% | 11% |
| D-12 | 0% | 0% | -5% | 0% | 7% | -9% | 22% | 13% | -16% | 16% | -1% | 18% | 11% | 21% | 5% | 1% |
| D-11 | 0% | 0% | 4% | 0% | -14% | -1% | -20% | -17% | -5% | -20% | -9% | -20% | -18% | -27% | -16% | -18% |
| D-10 | 0% | 0% | -9% | 0% | -10% | -1% | -8% | -9% | -3% | -6% | -5% | -21% | -8% | -8% | -6% | -4% |
| D-9 | 0% | 0% | -2% | 0% | 5% | -5% | 9% | 6% | 2% | 3% | 2% | 26% | 2% | 3% | 0% | -1% |
| D-8 | 0% | 0% | -1% | 0% | 4% | 4% | -3% | -4% | -1% | 8% | 10% | 10% | 10% | 8% | 7% | 11% |
| D-7 | 0% | 0% | 17% | 0% | 9% | 3% | 9% | 16% | 10% | 0% | -8% | 2% | 2% | 5% | 3% | -6% |
| D-6 | 0% | 0% | -14% | 0% | -11% | -2% | -5% | -9% | -7% | -1% | 4% | -23% | -6% | -3% | -3% | 8% |
| D-5 | 0% | 0% | 12% | 0% | 7% | 5% | 2% | 5% | 10% | 3% | 1% | 21% | 7% | -3% | 6% | 3% |
| D-4 | 0% | 0% | -5% | 0% | 5% | -6% | 19% | 20% | -1% | 8% | 3% | 19% | 8% | 27% | 8% | 3% |
| D-3 | 0% | 0% | 9% | 0% | 3% | 17% | -13% | -8% | 15% | -6% | 7% | -17% | 2% | -11% | 4% | 15% |
| D-2 | 0% | 0% | 5% | 0% | -2% | -15% | 10% | 11% | -7% | -2% | -8% | 7% | -4% | 7% | -8% | -15% |
| D-1 | 0% | 0% | -19% | 0% | -8% | -13% | 9% | -15% | -12% | 6% | -14% | -34% | -19% | -5% | -16% | -16% |
| D+1 | 0% | 0% | -47% | 0% | -57% | -66% | -68% | -77% | -70% | -79% | -90% | -72% | -80% | -85% | -89% | -93% |
| D+2 | 0% | 0% | 92% | 0% | 127% | 201% | 143% | 265% | 203% | 291% | 747% | 302% | 401% | 390% | 666% | 990% |
| D+3 | 0% | 0% | -19% | 0% | -13% | -2% | -14% | -21% | -14% | -5% | -3% | 6% | -18% | -2% | -6% | -6% |
| D+4 | 0% | 0% | 1% | 0% | 6% | 5% | 4% | -3% | 15% | 11% | 2% | -12% | 1% | 3% | -1% | 3% |
| D+5 | 0% | 0% | 13% | 0% | 5% | 12% | -1% | 22% | 24% | 2% | 22% | 29% | 26% | 5% | 29% | 26% |
| D+6 | 0% | 0% | -3% | 0% | -6% | -11% | -3% | -14% | -13% | 0% | -2% | -28% | -11% | -2% | -9% | 1% |
| D+7 | 0% | 0% | 5% | 0% | 7% | 6% | 13% | 15% | 10% | 3% | -1% | 19% | 13% | 9% | 7% | -2% |
| D+8 | 0% | 0% | 19% | 0% | 8% | 13% | 5% | 11% | 25% | 5% | -2% | 26% | 4% | 13% | 9% | 14% |
| D+9 | 0% | 0% | -4% | 0% | -7% | -8% | -3% | 12% | 2% | -4% | 5% | -14% | -15% | -6% | 1% | -2% |
| D+10 | 0% | 0% | 5% | 0% | 1% | 9% | -9% | -19% | -1% | -2% | 0% | -5% | 13% | -12% | -8% | -7% |
| D+11 | 0% | 0% | -11% | 0% | -9% | -10% | -7% | -13% | -11% | 0% | -6% | -25% | -18% | -4% | -10% | -2% |
| D+12 | 0% | 0% | 3% | 0% | 7% | 3% | 10% | 8% | 7% | -5% | -4% | 22% | 5% | 4% | -1% | -7% |
| D+13 | 0% | 0% | 3% | 0% | -2% | 4% | -7% | 7% | -3% | 4% | 4% | 5% | 9% | 6% | 11% | 11% |
| D+14 | 0% | 0% | -4% | 0% | 7% | -10% | 17% | 11% | -10% | 7% | 4% | 18% | 6% | 5% | 1% | -5% |
| D+15 | 0% | 0% | 12% | 0% | 6% | 18% | -9% | -3% | 19% | -2% | -1% | 10% | 14% | 6% | 6% | 12% |

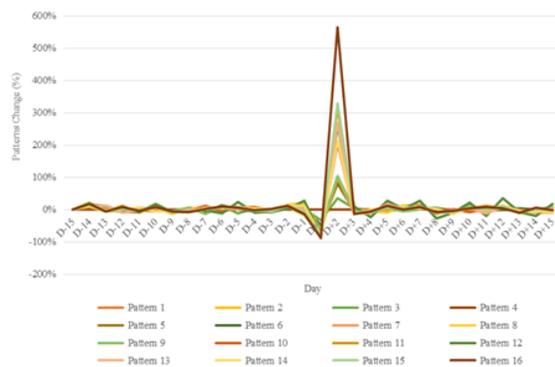

*Figure 15: Daily Motif Patterns Change 15 Days Before and After the 2014 Event*

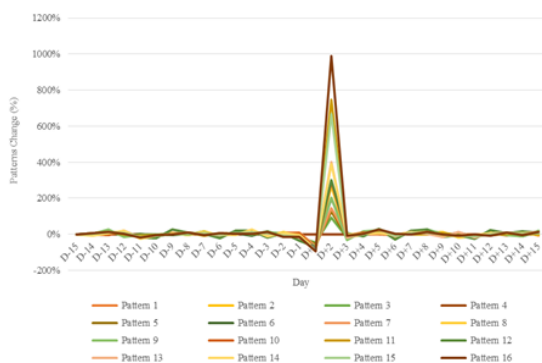

*Figure 16: Daily Motif Patterns Change 15 Days Before and After the 2015 Event*

Based on the table 4, table 5, figure 15, and figure 16, we found that the pattern 16 is the most disruptive pattern during the days of the event. This signify that this pattern is the most correlating triadic motif pattern with the event. This also means that this pattern qualifies to become the detector or early warning signal of the financial instability. Meanwhile, other motif patterns show less significant relation during the instability and less usable to be used as a detector of financial instability.

From the research result, we are able to model the financial instability condition based on learning to the religious event in Indonesia, where the banking transaction are normally heavily disrupted, the same way when the financial instability occur. The particular instability occurs on the day+1 of the event. We found the macro network structure measurement for 2014 and 2015 event indicates several phenomena explain in four point as follows:

- The number of nodes measurement that shows the number of banks transacting within the banking network reveals several banks are not perform any transaction in the day+1, thus indicate financial instability on that particular day.
- The number of edges which state the transaction activities conducted within the banking network. The finding reveals that several banks are reducing their transaction activities in the day +1.
- The network average distance indicates the average steps needed for a bank to transact with all the banks within the banking network. The finding reveals that the banks are having fewer shortcuts to transact with all other in day+1. The banking network are become sparser on that particular day.
- The network density measurement indicates the interaction intensity among banks within the banking network. The finding reveals that the banks have lower density, thus facing higher network breakage risk, since the network unable to tolerate the contagion default spreading.

The micro network structure measurement result indicates that there are triadic motif patterns shift within the network. This finding indicates the change of the banking transaction behavior. After exploring each network triadic motif pattern existence, the microstructure of banking network is instable in the day+1. We found pattern 16 is the most dominance pattern to follow the day+1 financial instability.

Pattern 16 consist of three reciprocal transaction between three banks. The construction of pattern 16 triadic motif reduces in day+1 and recovered in day+2. In the day+1. We have no proof to say that the pattern 16 transform to other triadic patterns, since there are no other increasing patterns when pattern 16 frequency decrease. The absence of several nodes is result in the triadic motif patterns decomposition to dyadic motif. The 3-nodes-sub-graph are split into 2-nodes-sub-graph. This finding reveals that during financial instability period, a bank is less connected to other two banks in reciprocal transaction activity.

Our work has enriched Squartini [21] works in several aspects: First, we propose network macro measurement to view the network macro condition, before going to micro measurement or network motif measurement. Second, we implement the idea to different banking network dataset, thus give different aspect, such as patterns dominance on different country. Third, we collect different insight regarding macro and micro behavior, such as the period of financial instability, thus it might give different implications to the bank regulations.

## 6. CONCLUSION AND FUTURE WORK

This study explores the banking network structural changes through macro and micro view. The Indonesian banking network dataset has been used to explain the network behavioral changes leading to financial instability. The major religious event is chosen to model the structural changes when the financial instability occurs. We have found that there are significant changes in macro and micro network measurement during the event. We also found which network motif pattern in micro network measurement to become the dominance one in correlating with disruptive event. We call this particular pattern as the financial instability detector.

Our finding brings new perspective on how we see the financial instability event. The current observations on global economic event, sometimes is hard to follow by most people without sufficiently proper knowledge to detect when the financial instability will occur. The exploration on banking network transaction gives complementary methodology for that purpose. This methodology is by far is straightforward and easier to follow, because it relies on simple object quantifications such as network properties and network motif.

There are three suggestions for the future work of this study. First is to generalize the methodology by represent the real-world complex financial instability sequence of event, not by simulating one-time event such as religious event, or political event. Second, in order to achieve the first suggestion, is by implementing the methodology to other banking network dataset, such as from other country. Third, by adopting dynamic measurement of network structure over time, which can enhance the parallel observations of real-time real-world event.